\documentclass{article}
\usepackage{graphicx}  
\usepackage{amsmath}   
\usepackage[compress]{cite}
\usepackage{amssymb}   
\usepackage{bm} 
\usepackage{dcolumn}
\usepackage{color}
\usepackage{mathrsfs}
\usepackage{amsfonts}
\usepackage{varioref}
\usepackage{textcomp}
\usepackage[parfill]{parskip}
\RequirePackage[colorlinks,citecolor=blue,urlcolor=magenta,linkcolor=blue]{hyperref}
\allowdisplaybreaks
\labelformat{section}{Section #1} 
\labelformat{subsection}{Section #1} 
\labelformat{subsubsection}{Section #1}
\labelformat{subsubsubsection}{Section #1}
\labelformat{equation}{Eq.~(#1)} 
\labelformat{figure}{Fig.~#1} 
\labelformat{subfigure}{Fig.~\thefigure#1} 
\labelformat{table}{Table.~#1} 
\labelformat{appendix}{Appendix #1}
\title{\bf Hilltop inflation and generation of helical magnetic field}

\author{Sumanta Chakraborty\footnote{sumantac.physics@gmail.com}$~^{1}$, 
Supratik Pal\footnote{supratik@isical.ac.in}$~^{2,3}$ and 
Soumitra SenGupta\footnote{tpssg@iacs.res.in}$~^{1}$\\
{$~^{1}$\small{School of Physical Sciences, Indian Association for the Cultivation of Science, Kolkata-700032, India}}\\
{$^{2}$\small{Physics and Applied Mathematics Unit, Indian Statistical Institute, Kolkata-700108, India}}\\
{$^{3}$\small{Technology Innovation Hub on Data Science, Big Data Analytics and Data Curation}}\\
{$^{\,\,\,\,\,\,}$\small{Indian Statistical Institute, Kolkata 700108, India}}}
\begin{document}
\maketitle
\begin{abstract}
Primordial magnetic field generated in the inflationary era can act as a viable source for the present day intergalactic magnetic field of sufficient strength. We present a fundamental origin for such a primordial generation of the magnetic field, namely through anomaly cancellation of $U(1)$ gauge field in quantum electrodynamics in the context of hilltop inflation. We have analysed at length the power spectrum of the magnetic field, thus generated, which turns out to be helical in nature. We have also found that magnetic power spectrum has significant scale-dependence giving rise to a non-trivial magnetic spectral index, a key feature of this model. Interestingly, there exists a large parameter space, where magnetic field of significant strength can be produced.
\end{abstract}
\section{Introduction and Motivation}\label{Mag_Intro}

Magnetic fields have made their presence felt at all the scales we have probed so far. Presence of magnetic fields at various length scales is a universal phenomenon as they cover a very broad scale, from planetary physics to galaxy clusters \cite{Grasso:2000wj,Beck:2000dc,Widrow:2002ud,Kandus:2010nw,Durrer:2013pga,Subramanian:2015lua}. There have been significant debate on the origin of this pervasive magnetic field at large scales. Broadly speaking, there exist two equally motivated pathways that one may follow to understand the origin of such large scale magnetic field. The first one advocates for an astrophysical origin, amplified later on by various mechanisms, e.g., dynamo mechanism \cite{Kulsrud:2007an,Brandenburg:2004jv,Subramanian:2015lua,Subramanian:2009fu}. While the other one ascribes the magnetic field to be of primordial origin, i.e., possible generation of magnetic field during the inflationary epoch \cite{Sharma:2017eps,Sharma:2018kgs,Jain:2012ga,Koley:2016jdw,Durrer:2010mq,Kanno:2009ei,Campanelli:2008kh,Martin:2007ue,Demozzi:2009fu,Bamba:2006ga,Ratra:1991bn,Ade:2015cva,Dolag:2010ni,Neronov:1900zz,Chowdhury:2018mhj,Vachaspati:1991nm,Turner:1987bw,Takahashi:2005nd,Agullo:2013tba,Ferreira:2013sqa,Atmjeet:2014cxa} or alternative scenarios like bouncing cosmology \cite{Chowdhury:2016aet,Chowdhury:2018blx}. Possible detection of magnetic field in voids may act as an acid test to favour one notion over the other.

Among all those possibilities, the primordial origin starting from inflation is much more appealing compared to others, primarily because of the elegance of the inflationary paradigm itself that gives rise to a natural mechanism of generating seeds of cosmological perturbations as well as its concordance with recent Cosmic Microwave Background (CMB) observations like WMAP 9 \cite{Hinshaw:2012aka}, Planck 2015  \cite{Ade:2015xua} and very recently, Planck 2018 \cite{Aghanim:2018eyx} data, among others. With future CMB missions like COrE \cite{Bouchet:2011ck}, LiteBIRD \cite{Matsumura:2013aja,Suzuki:2018cuy}, CMB-S4 \cite{Abazajian:2016yjj}, PRISM \cite{Andre:2013nfa}, PIXIE \cite{Kogut:2011xw} etc. chipping in, this is an exciting time to explore this particular possibility further and corroborate the results with those upcoming data.
 
However, in spite of all those appealing features,  the inflationary magnetogenesis is riddled with several difficulties. The most severe issue among them is how to generate the desired amount of  magnetic field strength that is consistent with the galactic scale results today \cite{Neronov:1900zz,Dolag:2010ni,Taylor:2011bn,Ade:2015cva}? Conformal invariance of any  electromagnetic action demands the magnetic field strength to decrease as $\sim 1/a^{2}$, which leads to a very feeble value for the magnetic field strength, that is clearly unable to account for the present day observations. This suggests that if one wishes to have a primordial origin for the magnetic field, conformal invariance of the electromagnetic action must be broken at the first place. As obvious, the usual Maxwell's equations in Friedmann-Robertson-Walker (FRW) background would not serve the purpose, as it leads to a  conformally invariant theory, see for example \cite{Widrow:2002ud,gravitation,Kandus:2010nw,Durrer:2013pga,Brandenburg:2004jv,Subramanian:2015lua}. Several possibilities to break the conformal invariance of the electromagnetic action have been invoked.  This includes any nontrivial coupling of the inflaton with the Maxwell's action. Several such models with interesting features are in vogue \cite{Sharma:2017eps,Sharma:2018kgs,Jain:2012ga,Koley:2016jdw,Durrer:2010mq,Kanno:2009ei,Campanelli:2008kh,Martin:2007ue,Demozzi:2009fu,Bamba:2006ga,Ratra:1991bn,Ade:2015cva,Dolag:2010ni,Neronov:1900zz,Chowdhury:2018mhj,Vachaspati:1991nm,Turner:1987bw,Takahashi:2005nd,Fujita:2012rb,Agullo:2013tba,Ferreira:2013sqa,Atmjeet:2014cxa,Caprini:2014mja,Kobayashi:2014sga,Atmjeet:2013yta,Fujita:2015iga,Campanelli:2015jfa,Tasinato:2014fia}. However, most of these models are plagued with at least one of the two drawbacks, namely the strong coupling and the back-reaction problem \cite{Demozzi:2009fu,Kobayashi:2014sga,Ferreira:2013sqa,Sharma:2017eps}. The strong coupling problem relates large values of the effective electric charge, rendering the effective field theory calculation invalid. On the other hand, if the strength of the magnetic field increases beyond the energy density of the background spacetime, effect of back-reaction cannot be neglected, resulting into the back-reaction problem. Another problematic avenue corresponds to presence of Schwinger effect \cite{Kobayashi:2014sga,Rajeev:2017smn,Sharma:2017ivh,Frob:2014zka,Stahl:2016geq,Stahl:2018idd}. If the strength of the electric field becomes strong enough there will be creation of particle anti-particle pair. These created particles being charged will increase the conductivity of the ambient medium, which in turn will reduce the magnetic field strength. Alternatively, in order to break conformal invariance one can also take shelter of nonlinear electrodynamics \cite{Campanelli:2007cg} or 3-form fields \cite{Koivisto:2011rm,Urban:2013aka}. However, these models take into account several assumptions and approximations.Alternatively, one may still work with conformally invariant Lagrangian, but which generates helical magnetic field. In contrast to non-helical fields, for helical magnetic fields power can be transferred from small to large scales through inverse cascading \cite{Brandenburg:2004jv,Campanelli:2007tc} and hence may generate magnetic field of sufficient strength. Presence of helical magnetic field may also have distinct signatures in the temperature anisotropy and polarizations of the CMB photons, providing interesting observational consequences \cite{Seto:2008sr}. There exists several other proposal to generate large scale magnetic field, e.g., the idea of resonant magnetic field \cite{Byrnes:2011aa}, inhomogeneous magnetic field \cite{Lewis:2004ef,Campanelli:2008kh} along with possibility of going beyond 4-dimensional electromagnetic theory \cite{Atmjeet:2013yta} are also around. In a completely different perspective, primordial magnetogenesis have also been realised to some extent in non-inflationary setup, say in the context of bouncing cosmology \cite{Chowdhury:2016aet,Koley:2016jdw,Chowdhury:2018blx}. Each of the proposals discussed above sound interesting and worth exploring further. 

While it is very much true that the investigation for a more suitable model should go on in the above-mentioned directions,  at the very outset, one notices that most of the models considered till date are phenomenological in nature. As the subject advances, one wonders if one can have a rather well-motivated framework of electromagnetic theory with strong theoretical background that can result in considerable magnetic field strength at the early universe. Herein lies the primary motivation of the present work. 

In this paper, we try to provide a consistent model of magnetogenesis from a fundamental perspective, namely, from a consistent quantum electromagnetic theory (for earlier attempts in this direction, see \cite{Field:1998hi,Anber:2006xt}). We also restrict ourselves to the good old 4-dimensional scenario and background inflationary setup. Besides obtaining the desired field strength for the magnetic field in the present epoch, it turns out that one can indeed avoid the above problems associated with the generation of primordial magnetic field. Thus we provide a completely new viewpoint to the inflationary magnetogenesis scenario, where we need not have to invoke any  ad-hoc coupling between the scalar field and  Maxwell's action  by hand, rather starting from a fundamental theory we have been able to demonstrate the possibility to arrive at a viable scenario. Interestingly enough, as it will turn out, the model has the potential to predict non-trivial features for the primordial magnetic field in general, which using future measurements can either reinforce the model or possibly rule it out. For example, we would  explore if one can have scale-dependent power spectrum for the magnetic field (along with the standard scale-invariant power spectrum), and if yes, what can be the possible constraints on the corresponding `magnetic spectral index', which would eventually be  a non-trivial parameter of the model under consideration.

The paper is organized as follows: In \ref{Mag_Sec_2}, we jot down basic field equations for the gravitation, scalar and electromagnetism, using that the scalar plus gravitational system has been solved within the context of hilltop inflation in \ref{hilltop}. We then engage ourselves in finding out the corresponding Maxwell's equations in FRW background keeping in mind inflationary evolution of the universe in \ref{Mag_Sec_3}. \ref{Mag_Sec_4} is entirely devoted to finding out the power spectra for electric and magnetic fields and subsequent solution of the vector potential. In \ref{Mag_Sec_5}, we analyse the strength of the magnetic field and explore if this scenario can generate sufficient magnetic field along with associated features therefrom.  We finally conclude with some open issues.

\textit{Notations and Conventions:} Throughout this paper we have set the fundamental constants $G$, $c$ and $\hbar$ to unity. We will work exclusively in the mostly positive signature. All the Greek letters are used to depict four dimensional coordinates, while Latin letters are used to present coordinates on the spatial sector.  
\section{Basic field equations: Setting up the stage}\label{Mag_Sec_2}

In this section, we will put forward the complete action for (gravity+matter) system, where the matter fields consist of the inflaton field $\phi$ and the vector potential $A_{\mu}$ associated with the electromagnetic field. We will assume that the electromagnetic field can be treated as a test field, while the background geometry and its evolution will be governed by the inflaton field $\phi$. The gravitational Lagrangian is taken to be the Ricci scalar and the inflaton field evolves in a potential $V(\phi)$, resembling hilltop inflationary scenario. On the other hand, for the electromagnetic field the Lagrangian will of course include the term $F_{\mu \nu}F^{\mu \nu}$. But for generation of a primordial magnetic field of significant strength, while maintaining conformal invariance, it is necessary to add an additional term in the Lagrangian, depicting the coupling between the electromagnetic and the inflaton field. Thus the complete action for this system takes the following form,
\begin{align}\label{Mag_v1_01}
\mathcal{A}&=\mathcal{A}_{\rm grav}+\mathcal{A}_{\phi}+\mathcal{A}_{\rm em}+\mathcal{A}_{\rm int}
\nonumber
\\
&=\frac{1}{16\pi G}\int d^{4}x\sqrt{-g}~R+\int d^{4}x\sqrt{-g}\left\{-\frac{1}{2}\partial _{\mu}\phi \partial ^{\mu}\phi-V(\phi)\right\}
\nonumber
\\
&\hskip 1 cm -\frac{1}{16\pi}\int d^{4}x\sqrt{-g}F_{\mu \nu}F^{\mu \nu}+\bar{\alpha}\int d^{4}x\sqrt{-g}\epsilon ^{\mu \nu \alpha \beta}\partial _{\mu}\phi A_{\nu}F_{\alpha \beta}~.
\end{align}
Here $\epsilon _{\mu \nu \alpha \beta}$ is the completely antisymmetric Levi-Civita tensor defined as, $\epsilon _{\mu \nu \alpha \beta}=\sqrt{-g}[\mu \nu \alpha \beta]$, where $[\mu \nu \alpha \beta]$ is the permutation symbol such that $[0123]=1$. The first term corresponds to the Einstein-Hilbert Lagrangian, while the second term is the action for the inflaton field. Among the terms in the last line, the first one is the canonical gauge invariant kinetic term for the electromagnetic field and the second term deals with the coupling between the electromagnetic vector potential and the inflaton field with coupling parameter $\bar{\alpha}$ (the dimensionless coupling constant being $\alpha M_{\rm Pl}$). 

Interestingly, the choice of the coupling term between electromagnetic field and the inflaton field is not ad-hoc and can be motivated along the following lines ---  During the inflationary epoch, one may expect the universe to inherit additional fundamental fields, e.g., the Kalb-Ramond field, whose presence leads to anomalies in the quantum domain, which can be removed, \'{a} la superstring theory, by a gauge Chern-Simons anomaly cancellation term \cite{Green:1987mn,Majumdar:1999jd} (also see \cite{Garretson:1992vt,Fujita:2015iga,Adshead:2015pva,Adshead:2016iae,Adshead:2018doq}). This results into a coupling between the abelian gauge field $A_{\mu}$ and the Kalb-Ramond field, which in four spacetime dimensions reduces to the interaction term considered in \ref{Mag_v1_01}. Moreover it turns out that the interaction term is indeed gauge invariant, since under the transformation, $A_{\mu}\rightarrow A_{\mu}+\partial _{\mu}\Psi$ and $\phi \rightarrow \phi$,
\begin{align}\label{Mag_v1_02}
\delta _{\Psi}\mathcal{A}_{\rm int}&=\bar{\alpha}\int d^{4}x\sqrt{-g}\epsilon ^{\mu \nu \alpha \beta}F_{\alpha \beta}\partial _{\mu}\phi \partial _{\nu}\Psi 
=\textrm{Total~Derivative}~,
\end{align}
since $\epsilon ^{\mu \nu \alpha \beta}$ is antisymmetric in the $(\mu,\nu)$ indices. Therefore, the Lagrangians before and after the gauge transformation differ by a total derivative term and hence they yield identical field equations \cite{gravitation}. 

The above action depends explicitly on three dynamical variables --- (a) the metric $g_{\mu \nu}$; (b) the vector potential $A_{\mu}$ and (c) the scalar field $\phi$. Thus we need to vary the action with respect to all of them. Variation with respect to the metric $g_{\mu \nu}$ will lead to the Einstein's equations, which read,
\begin{align}\label{Mag_v1_04}
G_{\alpha \beta}=8\pi G \left[\nabla _{\alpha}\phi \nabla _{\beta}\phi -\frac{1}{2}g_{\alpha \beta}\left\{\nabla _{\mu}\phi \nabla ^{\mu}\phi +2V(\phi)\right\}+\frac{1}{4\pi}F_{\mu \alpha}F^{\mu}_{~\beta}-\frac{1}{16\pi}g_{\alpha \beta}\left(F_{\mu \nu}F^{\mu \nu}\right)\right]~.
\end{align}
Here the first two terms in the energy momentum tensor, appearing in the right hand side of the above equation, arises from the scalar field. The rest of the terms in the energy momentum tensor are associated with the electromagnetic field. We observe that the coupling term between the scalar and the electromagnetic field does not contribute to the variation of the metric. This is because, in the action, the Levi-Civita tensor $\epsilon^{\mu \alpha \nu \beta}$ cancels the contribution from $\sqrt{-g}$. Moreover, from the energy momentum tensor presented in \ref{Mag_v1_04} it is obvious that the electromagnetic part is traceless and hence the coupling term between scalar and electromagnetic field respects conformal invariance.      

Proceeding further, we vary the action presented in \ref{Mag_v1_01} with respect to the gauge field $A_{\mu}$ to determine the modified Maxwell's equations, yielding,
\begin{align}\label{Mag_v1_09}
\nabla _{\mu}F^{\mu \nu}+8\pi \bar{\alpha}\epsilon ^{\mu \nu \alpha \beta}\nabla _{\mu}\phi F_{\alpha \beta}=0~.
\end{align}
Here we have invoked antisymmetry of the Levi-Civita tensor in order to write $\nabla _{\mu}A_{\beta}$ as $(1/2)F_{\mu \beta}$. Finally the dynamics of the scalar field gets determined by the following field equation,
\begin{align}\label{Mag_v1_12}
\square \phi &=\frac{\bar{\alpha}}{2}\epsilon ^{\mu \nu \alpha \beta}F_{\mu \nu}F_{\alpha \beta}
+\frac{\partial V}{\partial \phi}~.
\end{align}
Here, the d'Alembertian operator $\square$ stands for $g^{\mu \nu}\nabla_{\mu}\nabla_{\nu}$ and we have used the fact that $\epsilon ^{\mu \alpha \beta \rho}\nabla _{\alpha}F_{\beta \rho}=0$. Since, solving this set of coupled differential equations is difficult, we will assume that gravity is sourced by the scalar field $\phi$, which will generate an inflationary background, on which the ``test" electromagnetic field lives. In what follows, we will first describe the (gravity+scalar field) system, before taking up the case of ``test" electromagnetic field. 

\section{Hilltop inflation with quadratic potential}\label{hilltop}

In this section, we would like to explore the gravity plus scalar field system in a cosmological setting. The homogeneity and isotropy of the universe at a large scale, as well as the recent observational results advocating the universe to be spatially flat\footnote{To be precise the estimate of spatial curvature corresponds to $\Omega_k = 0.0007 \pm 0.0037$ from latest Planck 2018 data through TT, TE, EE + lowE + lensing + BAO data \cite{Aghanim:2018eyx}. Here TT means temperature temperature cross-correlation of CMB data, TE means cross-correlation between temperature and electric type polarization of CMB data and finally BAO stands for Baryon Acoustic Oscillation.} demands the following structure for the metric,
\begin{align}\label{Mag_v1_13}
ds^{2}=-dt^{2}+a^{2}(t)\left(dx^{2}+dy^{2}+dz^{2}\right)~.
\end{align}
The above metric ansatz has been expressed in terms of the cosmological time coordinate $t$, but for our purpose it will be advantageous to introduce a conformal time coordinate $\eta$, defined as $d\eta=dt/a(t)$ and hence the above line element can be recasted in the following form,
\begin{align}\label{Mag_v1_13N}
ds^{2}=a^{2}(\eta)\left(-d\eta^{2}+dx^{2}+dy^{2}+dz^{2}\right)~.
\end{align}
We will use this conformally flat form of the metric throughout our computations. Returning back to the dynamics of the scale factor, note that the gravitational field equations determine the evolution of the scale factor $a(\eta)$, with the energy-momentum tensor of the scalar field acting as the source. As a consequence, the isotropy and homogeneity of the universe demands the scalar field to be dependent on the conformal time alone, i.e., $\phi=\phi(\eta)$. We will treat this scalar field as the inflaton field and shall provide a hilltop inflationary scenario with a quadratic potential, given by $V(\phi)=V_{0}\left[1-\alpha(\phi-\phi_{0})^{2}\right]$ \cite{Boubekeur:2005zm,Bostan:2019wsd,Kohri:2007gq,Pal:2009sd}. Here the inflation starts at the maxima of the potential, when the scalar field has a value $\phi=\phi_{0}$. The coupled field equations for gravity and scalar field are best expressed in the following form,
\begin{align}
3\mathcal{H}^{2}&=M_{\rm Pl}^{-2}\left[\frac{1}{2}\phi'^{2}+a^{2}(\eta)V(\phi)\right]~,
\\
\phi''&+2\mathcal{H}\phi'+a^{2}(\eta)\left(\frac{\partial V}{\partial \phi}\right)=0~,
\end{align}
where, $\mathcal{H}=(a'/a)$. Thus the Hubble parameter under slow roll approximation (i.e., $\dot{\phi}^{2}\ll 1$) can be expressed in terms of the potential as,
\begin{align}
H=\frac{\mathcal{H}}{a(\eta)}=\sqrt{\frac{V_{0}}{3M_{\rm Pl}^{2}}}\left[1-\alpha\left(\phi-\phi_{0}\right)^{2}\right]^{1/2}
\end{align}
The slow-roll parameters associated with the above potential, or, equivalently with the Hubble parameter become,
\begin{align}
\epsilon_{H}&=2M_{\rm Pl}^{2}\left(\frac{1}{H}\frac{\partial H}{\partial \phi}\right)^{2}=\frac{2M_{\rm Pl}^{2}\alpha^{2}(\phi-\phi_{0})^{2}}{\left[1-\alpha\left(\phi-\phi_{0}\right)^{2}\right]^{2}}~,
\\
\eta_{H}&=2M_{\rm Pl}^{2}\left(\frac{1}{H}\frac{\partial^{2}H}{\partial \phi^{2}}\right)=-\frac{2M_{\rm Pl}^{2}\alpha}{\left[1-\alpha\left(\phi-\phi_{0}\right)^{2}\right]^{2}}~,
\\
\zeta_{H}&=4M_{\rm Pl}^{4}\left[\frac{1}{H^{2}}\frac{\partial H}{\partial \phi}\frac{\partial^{3} H}{\partial \phi^{3}}\right]=\frac{12M_{\rm Pl}^{4}\alpha^{3}\left(\phi-\phi_{0}\right)^{2}}{\left[1-\alpha\left(\phi-\phi_{0}\right)^{2}\right]^{2}}~.
\end{align}
So far, we have discussed the evolution equation for the scale factor, it is now time to consider the evolution of the scalar field as well. During the inflationary epoch, when slow roll conditions are satisfied, we may ignore the term $\ddot{\phi}$ in the evolution equation for the scalar field. The resulting field equation for the scalar field can be exactly solved, using the above potential and assuming a de Sitter background, yielding, $\phi-\phi_{0}\sim \eta^{-1}$, while the constraint on $\alpha$ will come from the Planck data. Thus for most of the duration of the inflationary epoch, as long as slow roll approximation holds, the scalar field evolves as, $\phi(\eta)=\phi_{0}-(b/\eta)$, where $b$ is a constant, to be fixed. This behaviour of the scalar field will play an important role in the subsequent discussion regarding the evolution of the electromagnetic field.  

Finally, let us briefly mention the consistency of the quadratic hilltop inflation model with the Planck data \cite{Boubekeur:2005zm,Ade:2015lrj}. For this purpose, it is worth mentioning the expressions for the tensor-to-scalar ratio $r$ and scalar spectral index $n_{\rm s}$ for the hilltop inflation model considered here. Under the slow roll approximation, these observables for the above quadratic hilltop inflation model are given by, $r\approx 32(\alpha M_{\rm pl}^{2})x^{2}(1-x^{2})^{-2}$ and $n_{s}-1=-4(\alpha M_{\rm pl}^{2})(1-x^{2})^{-1}-(3r/8)$ \cite{Ade:2015lrj}. Following \cite{Gron:2018rtj}, we may introduce the notation, $\delta_{\rm ns}=1-n_{\rm s}$, and inserting the expression for $r$, the last equation can be written as, $\delta_{\rm ns}=4\alpha M_{\rm pl}^{2}(1+2x^{2})(1-x^{2})^{-2}$. Hence, $\delta_{\rm ns}=\{8x^{2}/1+2x^{2}\}r$. Therefore, the dimensionless potential may be expressed in terms of the observables, as $x=\sqrt{(r/2)(4\delta_{\rm ns}-r)^{-1}}$. According to the Planck and BICEP2 data, $\delta_{\rm ns}=0.032$, $r=0.05$ giving, $x=0.57$. Here, $x\equiv \sqrt{\alpha}(\phi_{*}-\phi_{0})$, with $\phi_{*}$ being the scalar field associated with the pivot scale. As evident from \cite{Ade:2015lrj}, the above values of $r$ and $n_{\rm s}-1$ are consistent with the Planck data with $\log_{10}(1/\sqrt{\alpha}M_{\rm Pl})>1.02$. A similar analysis has also been performed in \cite{Gomes:2018uhv} for general hilltop models, where also the quadratic model was found to be consistent with the Planck data for $\sqrt{\alpha}M_{\rm Pl} \in [0,1]$. Thus for these choices of $\sqrt{\alpha}M_{\rm Pl}$, the quadratic model considered here is in consonance with the Planck data. This behaviour of the scalar field will be considered in the next section to determine the evolution of the electromagnetic field. 

\section{Modified Maxwell's Equations in Cosmological Background}\label{Mag_Sec_3}

In this section, we will determine the modified Maxwell's equations arising from the additional coupling between the inflaton and the electromagnetic field $A_{\mu}$. Unlike the scalar field, the electromagnetic field will be assumed to depend on both the time and the space coordinates. This does not affect the homogeneity and isotropy of the universe, since the energy density of the electromagnetic field is assumed to be not back-reacting on the background geometry. Thus given the background spacetime geometry, presented in \ref{Mag_v1_13}, the quantity $\nabla _{\mu}F^{\mu \nu}$  in \ref{Mag_v1_09} turns out to be,
\begin{align}\label{Mag_v1_14}
\nabla _{\mu}F^{\mu \nu}=\frac{1}{\sqrt{-g}}\partial _{\mu}\left(\sqrt{-g}F^{\mu \nu}\right)
=\frac{1}{a^{4}}\partial _{\mu}\left(a^{4}F^{\mu \nu}\right)~.
\end{align}
Further computation reveals that the zeroth component of the second term in \ref{Mag_v1_09} identically vanishes, while the spatial components are non-zero with the following expression,
\begin{align}\label{Mag_v1_18}
8\pi \bar{\alpha}\epsilon ^{\mu i\alpha \beta}\nabla _{\mu}\phi F_{\alpha \beta}
&=8\pi \bar{\alpha}\epsilon ^{0ijk}\phi'F_{jk}+16\pi \bar{\alpha}\epsilon ^{ji0k}\nabla _{j}\phi F_{0k}
\nonumber
\\
&=-8\pi \bar{\alpha}\left(\frac{1}{a^{4}}\right)\epsilon _{ijk}\phi'\left(\partial_{j}A_{k}-\partial_{k}A_{j}\right)
=-16\pi \bar{\alpha}\left(\frac{1}{a^{4}}\right)\epsilon _{ijk}\phi'\nabla_{j}A_{k}~.
\end{align}
where we have used the result, $\epsilon ^{\mu \nu \alpha \beta}=-(1/\sqrt{-g})[\mu \nu \alpha \beta]$, with $[\mu \nu \alpha \beta]$ being the completely antisymmetric permutation, with $[0123]=1$. Here as well all the spatial indices are to be raised/lowered using the flat space metric and `prime' denotes a derivative with respect to the conformal time $\eta$. Proceeding further, the zeroth component of the field equation for the vector potential in this isotropic and homogeneous background spacetime, reads,
\begin{align}\label{Mag_v1_19}
\partial _{\eta}\left(\partial _{i}A_{i}\right)-\partial _{i}\partial _{i}A_{0}=0~,
\end{align}
while the spatial component of the field equation for the vector potential reads,
\begin{align}\label{Mag_v1_20}
-\partial _{\eta}^{2}A_{i}+\partial _{\eta}\partial _{i}A_{0}+\left(\partial _{j}\partial _{j}A_{i}-\partial _{i}\partial _{j}A_{j}\right)
-16\pi \bar{\alpha}\epsilon _{ijk}\phi'\nabla_{j}A_{k}=0~.
\end{align}
Using the gauge freedom, the above equations can be further simplified, in particular, imposition of the Coloumb gauge condition (in which the zeroth component of the vector potential, i.e., $A_{0}$ identically vanishes) makes \ref{Mag_v1_19} trivially satisfied, provided $\partial _{i}A_{i}=0$. On the other hand, \ref{Mag_v1_20} becomes non-trivial and takes the following form,
\begin{align}\label{Mag_v1_22}
A''_{i}+16\pi \bar{\alpha}\epsilon _{ijk}\phi'\partial _{j}A_{k}-\partial _{j}\partial _{j}A_{i}=0~.
\end{align}
Thus use of the gauge invariance has indeed simplified the field equations for the vector potential to a large extent. Subsequently, we will analyse these equations to extract out information about the power spectrum and the field strength of the primordial electric and magnetic fields. 
\section{Power Spectra for Electric and Magnetic Fields}\label{Mag_Sec_4}

In this section, we will derive the power spectrum associated with both the electric and magnetic fields. The computation of the power spectrum requires two ingredients --- (a) first of all we must know the energy density associated with the electric and magnetic fields and the associated vacuum state, (b) secondly, we need to solve for the vector potential. We will first present the energy density of the electric and magnetic fields in terms of the vector potential and its derivatives. Subsequently we will solve for the vector potential, which when substituted in the above expressions for energy density of electric and magnetic fields, will provide the desired power spectrums.  

\subsection{Power spectra in terms of vector potential}

We start by analysing the various components of the matter energy momentum tensor $T_{\mu \nu}$. Since computation of energy density is our prime focus, the understanding of the $T^{0}_{0}$ (which is nothing but $-(1/a^{2})T_{00}$) component of the matter energy momentum tensor will suffice. Consequently, \ref{Mag_v1_04} results into the following expression for the $T^{0}_{0}$ component of the energy momentum tensor associated with the electromagnetic field,
\begin{align}\label{Mag_v1_31a}
T_{0}^{0}&=-\frac{1}{a^{2}}T_{00}=-\frac{1}{a^{2}}\Bigg\{\frac{1}{4\pi a^{2}}\left(A'_{i}\right)^{2}
+\frac{a^{2}}{16\pi}\left( -\frac{2}{a^{4}}\left(A'_{i}\right)^{2}+\frac{1}{a^{4}}F_{ij}F_{ij}\right)
-2a^{2}\bar{\alpha}\left(-\frac{1}{a^{4}}\epsilon _{ijk}\phi'A_{i}F_{jk}\right)
\nonumber
\\
&\qquad \qquad \qquad \qquad -\frac{2\bar{\alpha}}{a^{2}}\epsilon _{ijk}\dot{\phi}A_{i}F_{jk}\Big\}
=-\frac{1}{8\pi a^{4}}\left(A'_{i}\right)^{2}-\frac{1}{16\pi a^{4}}F_{ij}F_{ij}~.
\end{align}
We are now in a position to separate out the energy density associated with the electric and magnetic fields from the above expression for the energy density of the matter energy momentum tensor. The first term in \ref{Mag_v1_31a} is obviously the density of the electric field, while the other one depending only on the spatial derivatives of the vector potential contributes to the magnetic field. However, note that the notion of electric and magnetic fields are observer dependent. Given a Maxwell field tensor $F_{\mu \nu}$ one can construct the electric and magnetic field by projecting it appropriately using the four velocity of the desired observer, see e.g., \cite{gravitation,Thorne}. In this case the observer measuring the above electric and magnetic field corresponds to the fundamental observer moving with the cosmic flow. Hence the vacuum expectation value of the energy density associated with the electric field can be written as,
\begin{align}\label{Mag_v2_01}
\rho _{\rm E}=\langle 0|T^{0~\textrm{(E)}}_{0}|0\rangle
=-\frac{1}{8\pi a^{4}}\langle 0|A_{i}'^{2}|0\rangle~,
\end{align}
where $|0\rangle$ corresponds to the Bunch-Davies vacuum state associated with the initial electromagnetic field configuration. Note that, this vacuum expectation value is directly related to the power spectrum associated with electric and magnetic fields. This is because, the power spectrum is the Fourier transform of a two-point correlation function and the energy momentum tensor being quadratic in the field variable, encapsulates the information about the power spectrum. In a similar fashion, the energy density associated with the magnetic field can also be written down as
\begin{align}\label{Mag_v2_02}
\rho _{\rm B}=\langle 0|T^{0~\textrm{(B)}}_{0}|0\rangle
=\frac{1}{a^{4}}\langle 0|\left(-\frac{1}{16\pi}F_{ij}F_{ij}\right)|0\rangle~.
\end{align}
One then quantize the electromagnetic field by simply expanding it in the Fourier basis and writing in terms of the creation and annihilation operators, as: 
\begin{align}\label{Mag_v2_03}
A_{i}(\eta,\mathbf{x})=\sqrt{4\pi}\int \frac{d^{3}\mathbf{k}}{(2\pi)^{3}}
\sum _{p=1}^{2}\epsilon _{pi}
\left\{b_{(p)}(\mathbf{k})A_{p}(\mathbf{k},\eta)\exp(i\mathbf{k}.\mathbf{x})
+b_{(p)}^{\dagger}(\mathbf{k})A_{p}^{*}(\mathbf{k},\eta)\exp(-i\mathbf{k}.\mathbf{x})
\right\}~.
\end{align}
Here $\epsilon _{1i}$ and $\epsilon _{2i}$ are the polarization vectors in the standard linear polarization basis. However, due to presence of the Levi-Civita symbol in the Lagrangian it is instructive to rotate the linear polarization basis to the helicity basis, which is defined as $\epsilon _{\pm}^{i}=(1/\sqrt{2})(\epsilon _{1}^{i}\pm i \epsilon ^{i}_{2})$. For the rest of the discussion we will work with the helicity modes alone. The creation and annihilation operators associated with the vector potential are defined such that, $b_{(p)}(\mathbf{k})|0\rangle$ as well as its hermitian conjugate should vanish, where $|0\rangle$ is the Bunch-Davies vacuum state. Substituting the expression for the vector potential expressed in \ref{Mag_v2_03}, in the expectation values for energy density of the electric and magnetic field following \ref{Mag_v2_01} and \ref{Mag_v2_02}, we obtain
\begin{align}
\rho ^{\rm E}_{(p)}&=\frac{1}{2\pi ^{2}}\int dk~\frac{k^{2}}{a^{4}}|A_{p}'(k,\eta)|^{2}~;
\label{Mag_v2_08}
\\
\rho ^{\rm B}_{(p)}&=\frac{1}{2\pi ^{2}}\int dk~\left(\frac{k}{a}\right)^{4}|A_{p}(k,\eta)|^{2}~.
\label{Mag_v2_09}
\end{align}
Here we have used the fact that only the term involving $\langle 0|b_{(p)}(\mathbf{k})b_{(q)}^{\dagger}(\mathbf{k}')|0\rangle$ will contribute, as $b_{(p)}(\mathbf{k})|0\rangle$ and its hermitian conjugate identically vanishes. Further, the following identity involving sum over polarizations has also been used, 
\begin{align}\label{Mag_v2_06}
\sum _{p}\epsilon _{pi}\epsilon _{pk}=\delta _{ik}-\frac{1}{k^{2}}k_{i}k_{k}~.
\end{align}
Given the energy density associated with the electric and magnetic field as an integral over the momentum space, one can arrive at the following expressions for power spectra of electric and magnetic fields, namely
\begin{align}\label{Mag_v2_10}
\frac{d\rho^{\rm E}_{(p)}}{d\ln k}=\frac{1}{2\pi^{2}}\frac{k^{3}}{a^{4}}|A_{p}(k,\eta)'|^{2};\qquad 
\frac{d\rho^{\rm B}_{(p)}}{d\ln k}=\frac{k}{2\pi^{2}}\frac{k^{4}}{a^{4}}|A_{p}(k,\eta)|^{2}~.
\end{align}
As evident the power spectrum for the electric field depends on the time derivative of the vector potential, while the magnetic power spectrum is dependent on the vector potential alone. Further the $k$ dependence is also different in the two scenarios, in the electric power spectrum the dependence is through the $k^{3}$ term, while for magnetic power spectrum it is $k^{5}$. Hence it is very much likely, that when magnetic field power spectrum becomes scale invariant, the electric field power spectrum is not and vice versa. To understand the full dependence of the electric and magnetic power spectrum on the wave number, we need to solve for the electromagnetic vector potential. This is what we take up next.
\subsection{Solving for the vector potential}

In this section we would like to determine the vector potential $A(k,\eta)$, which is crucial in obtaining the power spectrum and its associated scale dependance. For this purpose, we need the evolution equation for the vector potential in terms of the conformal time, which has already been written down in \ref{Mag_v1_22}. That equation was written in the real space, but we need to rewrite it in Fourier space. This can be achieved by substituting the Fourier decomposition of the vector potential in \ref{Mag_v1_22}, which casts it in the following form
\begin{align}\label{Mag_v2_13}
\epsilon _{\pm i}A_{\pm}''(k,\eta)+16\pi \bar{\alpha}\epsilon _{ijk}\phi'(ik_{j})\epsilon _{\pm k}A_{\pm}(k,\eta)
-\epsilon _{\pm i}(ik_{j})(ik_{j})A_{\pm}(k,\eta)=0~.
\end{align}
As pointed out earlier, $\epsilon_{\pm i}$ corresponds to the polarization vectors associated with the helical modes of the electromagnetic wave (see also \ref{Mag_v2_03}). The above expression can be simplified further as the polarization vectors satisfy the following condition, namely, $\epsilon _{ijk}k_{j}\epsilon _{\pm k}=\mp i|\mathbf{k}|\epsilon _{\pm i}$. Broadly speaking, this relation holds as the photon polarization directions and the propagation direction of the electromagnetic waves are orthogonal. Thus the following differential equation for the Fourier component of the vector potential can be obtained,
\begin{align}\label{Mag_v2_14}
A''_{\pm}(k,\eta)+\left\{|\mathbf{k}|^{2}\pm \bar{\alpha}|\mathbf{k}|\phi'(\eta)\right\}A_{\pm}(k,\eta)=0~.
\end{align}
The above differential equation provides the final form of the differential equation for the vector potential, which one needs to solve to get the associated power spectrum. Note that depending on the helicity of the modes, the sign in front of $\bar{\alpha}$ changes. This explicitly demonstrates that the two helicity modes will evolve differently and hence will generate a non-zero magnetic helicity. For the moment being we will keep both the helicity modes in our analysis for completeness, but it will turn out that the negative helicity mode will have negligible influence on the physics we are interested in.

The first hindrance towards solving the differential equation for the vector potential, presented in \ref{Mag_v2_14} has to do with the fact that $\phi(\eta)$ is otherwise an arbitrary function of $\eta$ (remember this scalar field is, in general, an auxiliary field). Thus  the above equation for the vector potential can not give rise to a unique solution unless we specify some form of $\phi(\eta)$.  So, in order to keep life simple and to demonstrate that the mechanism works very well, we will concentrate on a particular situation when $\phi'(\eta)=b/\eta ^{2}$ with $b$ as an arbitrary constant. This is completely different from the scenario considered in \cite{Fujita:2015iga}, where the $\phi'$ term was taken to be $\sim \eta ^{-1}$, where a scale-invariant spectrum for the magnetic field may be obtained, but the electric field strength may exceed the inflationary energy scale, leading to back-reaction problem. The choice presented above, as we shall see later, will lead to scale dependent spectrum and will bypass the problem mentioned above.

In the standard inflationary setup, this is somewhat identical to considering a (quasi) de-Sitter evolution for which the Mukhanov-Sasaki equation has a term $a^{''}/a \sim -2/\eta^2$. Thus, our framework indeed is respecting inflationary evolution for the background, albeit the fact that $\phi$ is an auxiliary field here grants us the freedom of choice of its parameters, this feature will be revealed in due course. We, however, remind the reader that one can, in principle, make any other choice and redo the calculations afresh to check if that leads to a viable scenario. 

Under this choice the so-called {\it Mukhanov-Sasaki equation} for the vector potential takes the following form,
\begin{align}\label{Mag_v2_15}
A_{\pm}''(k,\eta)+\left\{|\mathbf{k}|^{2}\pm\frac{\bar{\alpha}|\mathbf{k}|b}{\eta ^{2}}\right\}A_{\pm}(k,\eta)=0~.
\end{align}
In order to solve the above equation, let us now introduce a new variable  $x=-k\eta$ (here and in what follows we denote $k=|\mathbf{k}|$), along with a new vector potential $\mathcal{A}_{\pm}(x)=(A_{\pm}/\sqrt{x})$, such that the above  equation takes the form, 
\begin{align}\label{Mag_v2_17}
x^{2}\frac{d^{2}\mathcal{A}_{\pm}}{dx^{2}}+x\frac{d\mathcal{A}_{\pm}}{dx}+\left\{x^{2}-\left(\frac{1}{4}\mp\bar{\alpha}kb \right)\right\} \mathcal{A}=0~.
\end{align}
Note that in order to arrive at \ref{Mag_v2_17} we have also multiplied the differential equation throughout by $x^{3/2}$. As evident, the above differential equation is a Bessel's differential equation with the following well-known solution,
\begin{align}\label{Mag_v2_18}
\mathcal{A}_{\pm}(x)=C_{1,\pm}J_{\nu_{\pm}}(x)+C_{2,\pm}Y_{\nu_{\pm}}(x)~;\qquad 
\nu_{\pm}^{2} =\frac{1}{4}\left(1\mp4\bar{\alpha}kb\right)~,
\end{align}
where $C_{1,\pm}$ and $C_{2,\pm}$ are arbitrary constants, $J_{\nu_{\pm}}$ is the Bessel's function for first kind and $Y_{\nu_{\pm}}$ is the modified Bessel's function. At this stage it is worth emphasizing that unlike the normal scenarios, in the present context, the index of the Bessel function itself, namely $\nu_{\pm}$ depends on the wave number $k$. Therefore, returning back to the conformal time coordinate and the original dynamical variable $A_{\pm}$ from $\mathcal{A}_{\pm}$, the solution for the vector potential takes the following form,
\begin{align}\label{Mag_v2_19}
A_{\pm}(k,\eta)=\sqrt{-k\eta}\Big\{C_{1,\pm}J_{\nu_{\pm}}(-k\eta)+C_{2,\pm}Y_{\nu_{\pm}}(-k\eta)\Big\}~.
\end{align}
However rather than using the modified Bessel function it is advantageous to replace the modified Bessel function $Y_{\nu_{\pm}}$ to Bessel functions $J_{\nu_{\pm}}$ and $J_{-\nu_{\pm}}$ with appropriate coefficients, such that the above solution for the vector potential involving both $J_{\nu_{\pm}}$ and $Y_{\nu_{\pm}}$ can be rewritten as, 
\begin{align}\label{Mag_v2_21}
A_{\pm}(k,\eta)=\sqrt{-k\eta}\Big\{D_{1,\pm}J_{\nu_{\pm}}(-k\eta)+D_{2,\pm}J_{-\nu_{\pm}}(-k\eta)\Big\}~.
\end{align}
Here we have introduced a new set of arbitrary constants $D_{1,\pm}$ and $D_{2,\pm}$, which are algebraically related to the original ones. Finally in order to determine the constants $D_{1,\pm}$ and $D_{2,\pm}$ we need to use appropriate boundary conditions, which correspond to super and sub-horizon scales.  First of all consider the limit $k\eta \rightarrow 0$, this in turn implies $k^{-1}$ becoming larger, i.e., the modes are in the super-horizon scale, much larger than the Hubble radius. In this case we can use the power law expansion of the Bessel function, leading to,
\begin{align}\label{Mag_v2_24}
A_{\pm}(k,\eta)|_{k\eta \rightarrow 0}=\sqrt{-k\eta}\left\{\frac{D_{1,\pm}}{2^{\nu_{\pm}}\Gamma(\nu_{\pm}+1)}\left(-k\eta\right)^{\nu_{\pm}}+\frac{D_{2,\pm}}{2^{-\nu_{\pm}}\Gamma(-\nu_{\pm}+1)}\left(-k\eta\right)^{-\nu_{\pm}}\right\}~.
\end{align}
Similarly, for modes living deep within the Hubble horizon we have $k^{-1}$ to be small, which in turn implies a large $k\eta$ limit. In this context the Bessel's function can be expanded in terms of Sines and Cosines with appropriate argument. These expansions when substituted in the expression for the vector potential presented in \ref{Mag_v2_21} yields,
\begin{align}\label{Mag_v2_26}
A_{\pm}(k,\eta)|_{k\eta \rightarrow \infty}=\sqrt{\frac{2}{\pi}}\left[D_{1,\pm}\cos \left(-k\eta-\frac{\pi}{2}\left\{\nu_{\pm}+\frac{1}{2}\right\} \right)+D_{2,\pm}\sin \left(-k\eta+\frac{\pi}{2}\left\{\nu_{\pm}+\frac{1}{2}\right\} \right)\right]~.
\end{align}
Rather than writing in terns of sinusoidal functions it is also possible to expand out the vector potential in terms of $\exp(\pm ik\eta)$. This is advantageous because, the initial condition for the vector potential has to do with the fact that the vector potential should be made out of positive frequency mode functions alone, deep inside the Hubble radius, such that,
\begin{align}\label{Mag_v2_28}
A(k,\eta)|_{k\eta \rightarrow -\infty}=\frac{1}{\sqrt{2k}}e^{-ik\eta}~.
\end{align}
Thus imposing this condition on the expansion of the vector potential presented in \ref{Mag_v2_26}, we can determine the respective expressions for the unknown coefficients $D_{1,\pm}$ and $D_{2,\pm}$. Therefore, finally plugging back all the expressions for the unknown coefficients, the vector potential in the super-horizon scale takes the following form,
\begin{align}\label{Mag_v2_31}
A_{\pm}(k,\eta)= \frac{1}{\sqrt{k}}\left\{A_{1,\pm}\left(-k\eta\right)^{\nu_{\pm}+(1/2)}+A_{2,\pm}\left(-k\eta\right)^{-\nu_{\pm}+(1/2)}\right\}~,
\end{align}
where the constants $A_{1,\pm}$ and $A_{2,\pm}$ have been introduced for convenience, having the following definitions,
\begin{align}\label{Mag_v2_32}
A_{1,\pm}=\frac{\sqrt{\pi} e^{-i(\pi/2)\{\nu_{\pm}+(1/2)\}}}{2^{\nu_{\pm}+1}\cos \left(\pi\nu_{\pm}+\{\pi/2\} \right)}\frac{1}{\Gamma(\nu_{\pm}+1)};
\qquad
A_{2,\pm}=\frac{\sqrt{\pi} e^{i(\pi/2)\{\nu_{\pm}+(3/2)\}}}{2^{1-\nu_{\pm}}\cos \left(\pi\nu_{\pm}+\{\pi/2\} \right)}\frac{1}{\Gamma(-\nu_{\pm}+1)}~.
\end{align}
An important point to note over here is that even though it appears that the above constants are independent of $k$, such is not the case. The dependence on the wave number is through the quantity $\nu_{\pm}$ appearing in the above expression, which corresponds to, $\nu_{\pm}^{2}=(1/4)\left(1\mp4\bar{\alpha}kb\right)$. This is not surprising as the term $\nu_{\pm}$ is related to the corresponding spectral index $n_B$ that takes care of the  scale-dependence of the magnetic power spectrum, if any. We will elaborate on this in the next section.

Before proceeding further, let us settle the issue of helicity associated with the electromagnetic modes. So far we have kept both the positive and negative helicity modes in our computation, but as evident from the expression for $\nu_{\pm}$, these modes will behave very differently depending on the parameter space of the combination $\bar{\alpha}|\mathbf{k}|b$, in particular, the sign of the parameter $b$ will be crucial. The measure of helicity associated with the electromagnetic field modes corresponds to the combination $\{|A_{+}|^{2}-|A_{-}|^{2}\}a^{-2}$. Thus for small $|k\eta|$, unless the magnitude of the respective vector potential grows faster, it will have very little influence on the helicity at the super-horizon scale. For negative values of $\bar{\alpha}b$, it immediately follows that $\nu_{+}$ will always be real while $\nu_{-}$ will be imaginary and hence at super-horizon scales influence from the negative helicity modes will be vanishingly small. Further since we expect the energy density of the $\phi$ field to decreases with time (so that it does not affect the cosmological dynamics at late stages) it is important that $\phi'$ and hence the parameter $b$ is negative. Thus under the assumption of small back-reaction by the scalar field the negative helicity modes remain sub-dominant compared to the positive helicity mode in the super-horizon scale. This allows us to concentrate on the positive helicity modes for all intents and purposes.
\section{Strength of The Magnetic Field and its Scale Dependance}\label{Mag_Sec_5}

Having derived the vector potential, let us now assess the strength of the electromagnetic field, in particular that of the magnetic field. This requires the knowledge of the solution for the vector potential derived above, which has two independent branches. The strength and power spectrum of the magnetic field will depend on whichever of the two branches dominate, as the other one can be safely neglected. This in turn depends on the sign of the parameter $\nu$. Thus for positive and negative choices of $\nu$ the strength as well as the power spectrum of the magnetic field will be different as we pick up the dominating solution and neglect the sub-dominant one. For $\nu>0$, it is evident that $\{\nu+(1/2)\}>\{-\nu+(1/2)\}$ and hence in the super-horizon scale the second term in \ref{Mag_v2_31} dominates, while for $\nu<0$, the situation is just the opposite. Consequently, from \ref{Mag_v2_10} we have the following power spectrum for the magnetic field,
\begin{align}
\frac{d\rho _{\rm B}}{d\ln k}&=\frac{|A_{2}|^{2}}{2\pi ^{2}}H^{4}\left(-k\eta \right)^{5-2\nu};\qquad \nu>0,~(k/a)\simeq -Hk\eta~;
\label{Mag_v2_35a}
\\
\frac{d\rho _{\rm B}}{d\ln k}&=\frac{|A_{1}|^{2}}{2\pi ^{2}}H^{4}\left(-k\eta \right)^{5+2\nu};\qquad \nu<0,~(k/a)\simeq -Hk\eta~.
\label{Mag_v2_35b}
\end{align}
On the other hand, the determination of the strength of the electric field will require the derivative of the vector potential as well, which can be determined straightaway from \ref{Mag_v2_31}. This leads to the following expression for the power spectrum of the electric field,
\begin{align}
\frac{d\rho _{\rm E}}{d\ln k}&=\frac{|A_{2}|^{2}}{2\pi ^{2}}\left(-\nu+\frac{1}{2}\right)^{2}
H^{4}\left(-k\eta \right)^{3-2\nu};\qquad \nu>0,~(k/a)\simeq -Hk\eta~;
\label{Mag_v2_36a}
\\
\frac{d\rho _{\rm E}}{d\ln k}&=\frac{|A_{1}|^{2}}{2\pi ^{2}}\left(\nu+\frac{1}{2}\right)^{2}
H^{4}\left(-k\eta \right)^{3+2\nu};\qquad \nu<0,~(k/a)\simeq -Hk\eta~.
\label{Mag_v2_36b}
\end{align}
Note that in plain sight it may appear that whenever $\nu=5/2$ (or, $\nu=-5/2$), the magnetic power spectrum for positive $\nu$ (or, negative $\nu$) is scale invariant. However, the parameter $\nu$ itself depends on the pivot scale $k$ and hence the power spectrum is never scale invariant. Thus in this particular model, in contrast to some other scenarios (see e.g., \cite{Fujita:2015iga}), it is not possible to make the electric and magnetic power spectrum scale invariant i.e., independent of the pivot scale of measurement. Consequently, one can define a magnetic  spectral index $n_B$ for different values of $\nu$ (and hence of $\bar \alpha b$),  which will act as a non-trivial parameter, along with the magnetic power spectrum, under consideration \footnote{In principle, one can also define an electric spectral index with similar arguments, but as it will turn out, the electric field strength is not going to play significant role here and hence, we will eventually ignore it.}. \emph{This acts as a distinct signature of our model}. 
\begin{figure*}
\begin{center}

\includegraphics[scale=0.5]{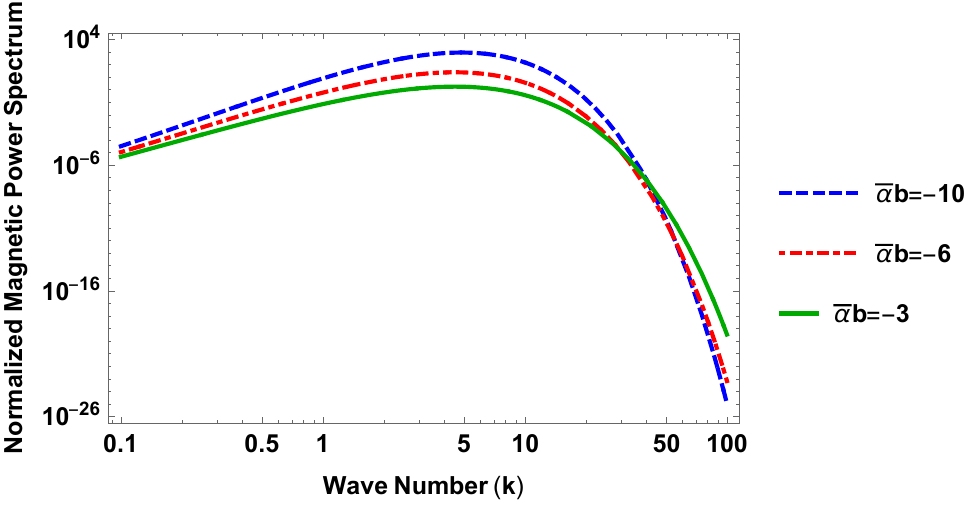}~~
\includegraphics[scale=0.5]{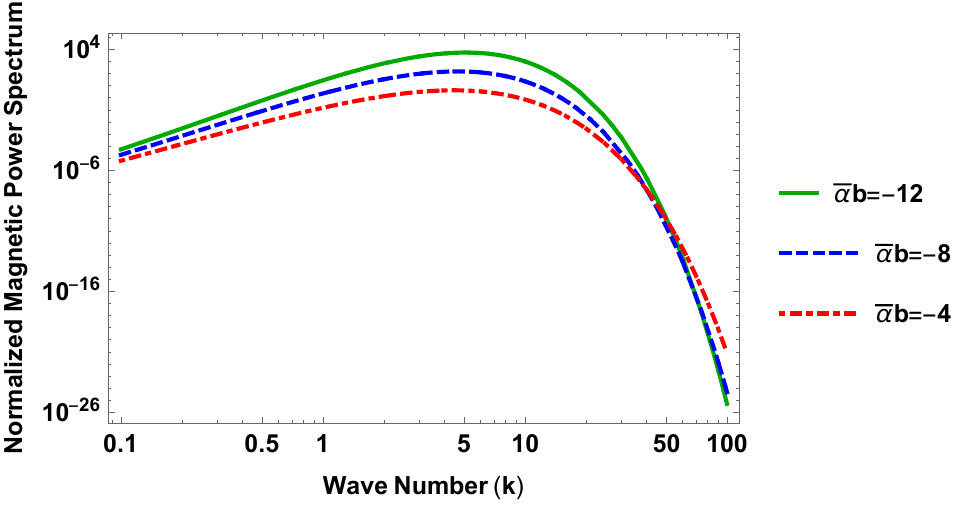}\\

\caption{Scale dependance of the magnetic power spectrum has been demonstrated. The figure on the left demonstrates normalized magnetic power spectrum for negative values of $\nu$, while that on the right demonstrates the same for positive choices of $\nu$. As evident, for smaller values of $|\bar{\alpha}b|$ the scale dependance of the power spectrum weaker. See, text for more discussion.}\label{fig_01}
\end{center}
\end{figure*}

\begin{figure*}
\begin{center}

\includegraphics[scale=0.5]{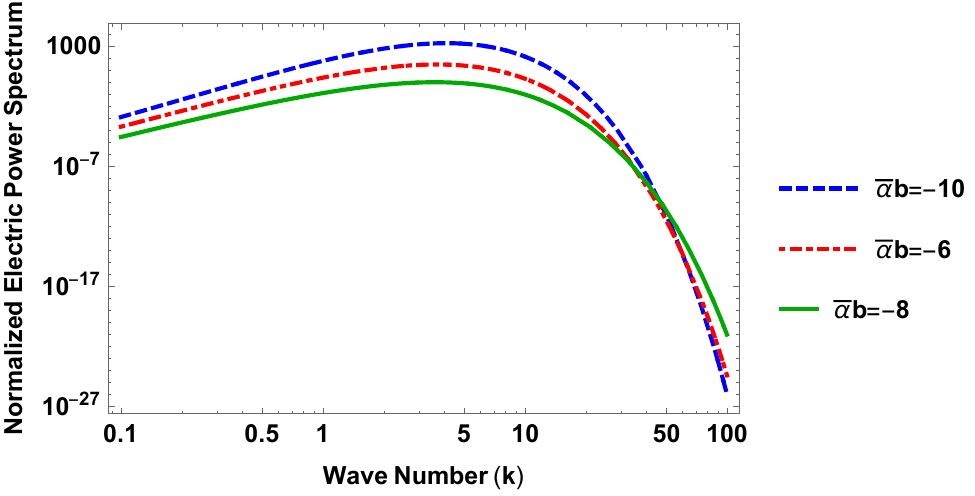}~~
\includegraphics[scale=0.5]{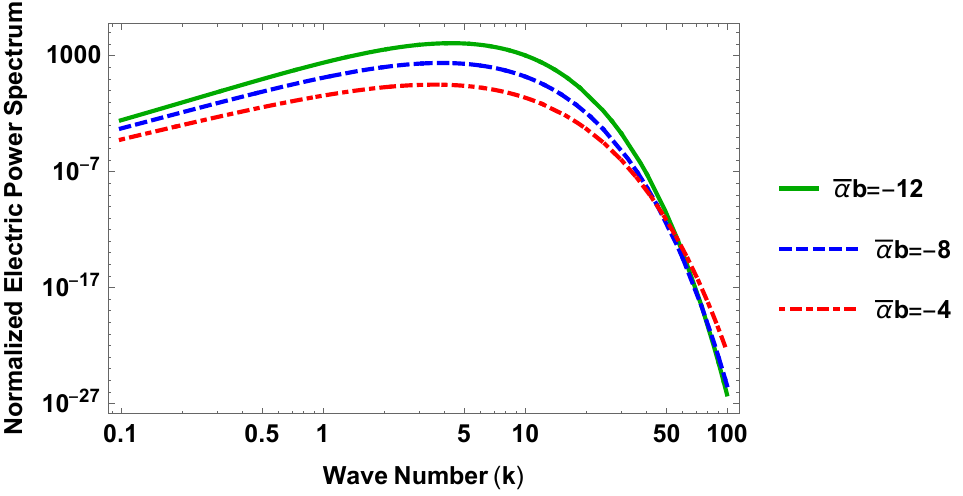}\\

\caption{The above figures demonstrate scale dependance of electric power spectrum. The figure on the left demonstrates the normalized electric power spectrum for negative choices of $\nu$, while that on the right demonstrates the same for positive values $\nu$. Alike the magnetic power spectrum, these figures also depict a very similar scale dependant nature of electric power spectrum.}\label{fig_02}
\end{center}
\end{figure*}
 
This feature has been explicitly demonstrated in \ref{fig_01} and \ref{fig_02} respectively. In particular \ref{fig_01} depicts the behaviour of magnetic power spectrum, normalized to some appropriate scale by $H^{4}$ with the wave number $k$ for both negative and positive choices of $\nu$. As evident, the scale dependence of the magnetic power spectrum is manifest in the plots. Only for smaller values of the quantity $\bar{\alpha}b$, the scale dependance is much weaker. A very much similar feature is maintained in the power spectrum for electric field as well. Hence scale dependance of power spectrum is a distinct signature of this model. 
 
Let us now concentrate on the strength of the magnetic field in the present epoch, i.e., it is important to know whether the model can generate magnetic field of sufficient strength. For that purpose we need to know the quantity $(\sigma/H)$, where $\sigma$ is the conductivity of the Plasma and $H$ is the associated Hubble parameter, both during the inflationary epoch and immediately after the same. During inflation the universe was a very poor electrical conductor \cite{Barrow:2011ic} and hence we need not worry much about the magneto-hydrodynamics of the plasma, as electric currents were small. However after inflation the universe becomes a very good conductor and hence the electric currents become important. This can be understood along the following lines: One starts with the collision time scale $\tau_{\rm c}$ for a charged particle moving in that plasma and determines the terminal velocity associated therewith. If the temperature of the plasma is $T$, then the scattering cross-section turns out to be, $\sigma \sim (ne^{2}\tau_{\rm c}/T)$, where $n$ is the number density of charged particles with charge $e$ and we have set the Boltzmann's constant $k_{B}=1$. The determination of $\tau_{\rm c}$ follows from the scattering of the charge particle with another one. This essentially leads to,
\begin{align}\label{Mag_v2_38}
\sigma \sim \frac{T}{\alpha \ln \Lambda}~,
\end{align}
where $\alpha \sim e^{2}$ is sort of a ``fine structure constant" and $\Lambda$ is a parameter depending on the details of the scattering. At the end of the inflation, we assume instant pre-heating, i.e., the universe makes a sudden jump from de-Sitter phase to radiation dominated universe. Then in the radiation dominated universe, we have one of the Einstein's equations to read, $t=H^{-1}=(m_{\rm pl}/T^{2})$, where $m_{\rm pl}^{2}$ is the inverse of the Newton's gravitational constant. Therefore using \ref{Mag_v2_38} we obtain, $(\sigma/H)\sim (m_{\rm pl}/T)\gg 1$. 

Thus after the inflationary epoch due to high electrical conductivity of the Plasma we can write $J^{i}=\sigma E^{i}$, where $E^{i}$ are the components of the electric field. Then the corresponding vector potential have two solutions, one independent of time and the other behaving as $\exp(-4\pi \sigma t)$. Since $\sigma t\sim (\sigma/H)$, it follows that the exponential will be vanishingly small. Thus the vector potential remains constant with time. This suggests that the electric field has very little contribution, while the magnetic field is the dominant piece. Hence one can use the information that universe was a very good conductor post inflation to argue that the energy density of the magnetic field in the present epoch must be related to that at the end of the inflation through the relation, $\{\rho _{\rm B}(0)/\rho_{\rm B}(t_{\rm f})\}=\{a_{\rm f}/a_{0}\}^{4}$. Here $t_{\rm f}$ denotes the time scale depicting the end of inflation and $a_{\rm f}$ is the scale factor at the end of the inflation. To determine the ratio of the scale factors one may use conservation of entropy, leading to, 
\begin{align}\label{Mag_v2_40}
\frac{a_{0}}{a_{f}}=\left(\frac{g_{\rm f}}{g_{0}}\right)^{1/3}\frac{T_{\rm f}}{T_{0}}~,
\end{align}
where $g$ stands for the average number of particle species present at a certain epoch in the universe. Thus using this expression and writing down the temperature in terms of Hubble parameter using the Einstein's equations, we obtain the ratio of the scale factors
\begin{align}\label{Mag_v2_41}
\frac{a_{0}}{a_{f}}=0.9\times 10^{29}\left(\frac{H}{10^{-5}M_{\textrm{pl}}}\right)^{1/2}~.
\end{align}
Finally, to get an estimate for the strength of the magnetic field we choose $\nu=(5/2)$ and $k\eta \sim -1$, which yields $A_{2}=-(3/\sqrt{2})$ and $\bar{\alpha}b\sim -6$. Then using the fact that energy density of the electromagnetic field falls of as $a^{-4}$, we can relate the strength of the magnetic field at the end of inflation to the present day strength of the same using \ref{Mag_v2_41}. This results into the following strength for the magnetic field in the present epoch, given the above parameter values, namely
\begin{align}\label{Mag_v2_42}
B_{0}\sim 0.5\times 10^{-10}~\textrm{Gauss}~.
\end{align}
The above result gives us a typical value for the magnetic field strength in the present epoch as obtained from our framework. However, this choice of $\nu$ apparently gives rise to only scale-invariant power spectrum for the magnetic field, as apparent from the discussions after \ref{Mag_v2_35a}. However, since $\nu$ depends on $k$, the power spectrum or strength so obtained is  actually scale dependent. We will comment on this later on.

As argued earlier, one can, however, arrive at a scale-dependent power spectrum in the present scenario for any other value of $\nu \neq (5/2)$,  resulting in a magnetic spectral index $n_B \neq 1$. This can be achieved by exploiting the freedom of choice of the parameter $\nu$ (by choosing the free parameter $\bar{\alpha}b$ in our theory), which is also scale dependent. This is interesting in particular, since Planck 2015 \cite{Ade:2015xua} and the following Planck 2018 \cite{Aghanim:2018eyx} results show a signature of scale-dependent power spectrum, at least for the scalar perturbations, at 5-$\sigma$. So, in principle, its magnetic counterpart may also have some scale-dependence, and if so, that needs to be explored. 
\begin{figure*}[htb]
\begin{center}

\includegraphics[scale=0.8]{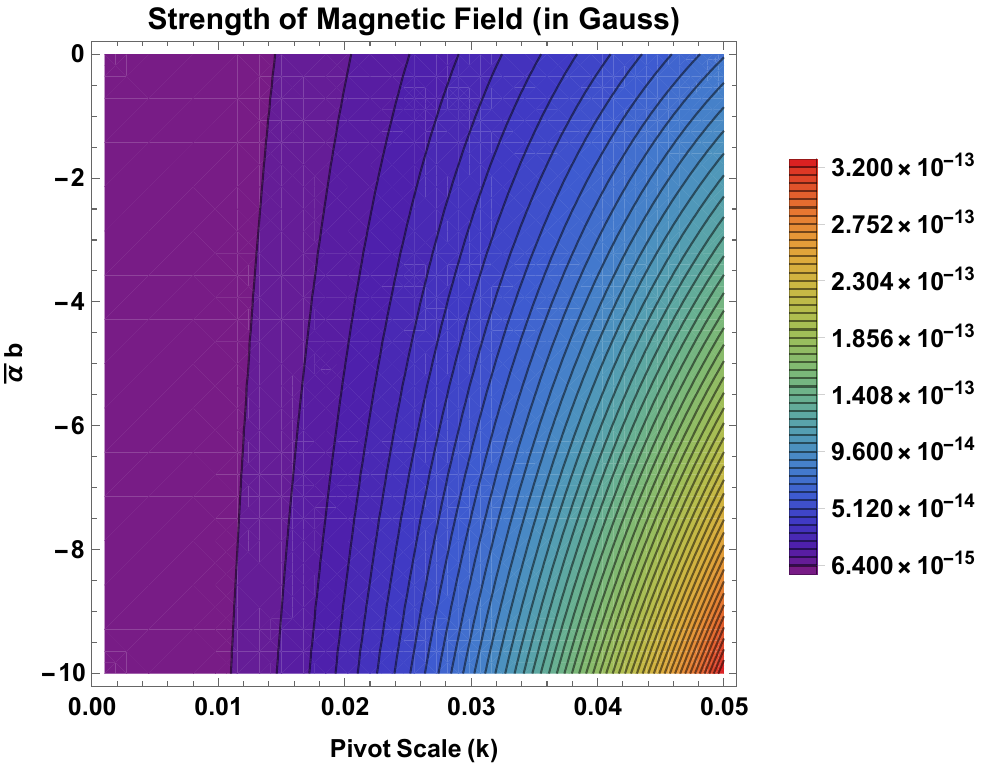}

\caption{The above figure demonstrates the magnetic field strength in Gauss in the present epoch with positive values for $\nu$ (determined by the parameter $|\bar{\alpha}b|$), with both the pivot scale $k$ (in Mpc$^{-1}$) and the parameter $\bar{\alpha}b$. Each contour depicts one constant value of the magnetic field. As evident from the colour coding the strength of the magnetic field varies from $B_{0}\sim 10^{-13}$ Gauss for large $k$ and $|\bar{\alpha}b|$ to $B_{0}\sim 10^{-15}$ Gauss for small $k$ and small $|\bar{\alpha}b|$.}\label{fig_03}
\end{center}
\end{figure*}

\begin{table*}
\begin{center}
\caption{The numerical estimations of present day magnetic field strength in Gauss is being presented for different choices of the parameter $\bar{\alpha}b$ for five different choices of the pivot scale $k$ (in Mpc$^{-1}$). As evident for smaller values of the parameter $|\bar{\alpha}b|$ as well as choice of the pivot scale, the magnetic field strength decreases. See text for more discussion.}
\label{Table_01}       
%
%
\begin{tabular}{p{0.5cm}p{2.5cm}p{2.5cm}p{2.5cm}p{2.5cm}}
\noalign{\smallskip}
\hline\noalign{\smallskip}
\hline\noalign{\smallskip}
        			&    			                		        &     Strength of         &                                         & \\
			&							& the				&					&\\
        			 &                           	     			        &	Magnetic Field		&                                          & \\
\hline\noalign{\smallskip}
\hline\noalign{\smallskip}
$|\bar{\alpha}b|$   &  (Pivot scale                   &  (Pivot scale                   & (Pivot scale                      & (Pivot scale      \\
			    &   ~~~~k=0.09)	                  &   ~~~~k=0.05)                &  ~~~~k=0.01)                   &   ~~~~k=0.005)\\
\hline \noalign{\smallskip}
\hline \noalign{\smallskip}
1                           & $1.591 \times 10^{-13}$   & $4.571 \times 10^{-14}$   & $1.779\times 10^{-15}$     & $4.382\times 10^{-16}$ \\
2                           & $1.831 \times 10^{-13}$   & $5.264 \times 10^{-14}$   & $1.844\times 10^{-15}$     & $4.478\times 10^{-16}$ \\
3                           & $2.108 \times 10^{-13}$   & $5.830 \times 10^{-14}$   & $1.911\times 10^{-15}$     & $4.575\times 10^{-16}$ \\
4                           & $2.423 \times 10^{-13}$   & $6.449 \times 10^{-14}$   & $1.981\times 10^{-15}$     & $4.674 \times 10^{-16}$ \\
5                           & $2.779 \times 10^{-13}$   & $7.124 \times 10^{-14}$   & $2.052\times 10^{-15}$     & $4.774\times 10^{-16}$ \\
6                           & $3.179 \times 10^{-13}$   & $7.858 \times 10^{-14}$   & $2.125\times 10^{-15}$     & $4.876\times 10^{-16}$ \\
7                           & $3.628 \times 10^{-13}$   & $8.655 \times 10^{-14}$   & $2.201\times 10^{-15}$     & $4.980\times 10^{-16}$ \\
8                           & $4.129 \times 10^{-13}$   & $9.519 \times 10^{-14}$   & $2.278\times 10^{-15}$     & $5.086\times 10^{-16}$ \\ 
9                           & $4.688 \times 10^{-13}$   & $1.045 \times 10^{-13}$   & $2.358\times 10^{-15}$     & $5.193\times 10^{-16}$ \\
10                         & $5.310 \times 10^{-13}$   & $1.146 \times 10^{-13}$   & $2.441\times 10^{-15}$     & $5.303\times 10^{-16}$ \\
\noalign{\smallskip}
\hline\noalign{\smallskip}
\hline \noalign{\smallskip}
\end{tabular}
\end{center}
\end{table*}

\begin{figure*}[htb]
\begin{center}

\includegraphics[scale=0.8]{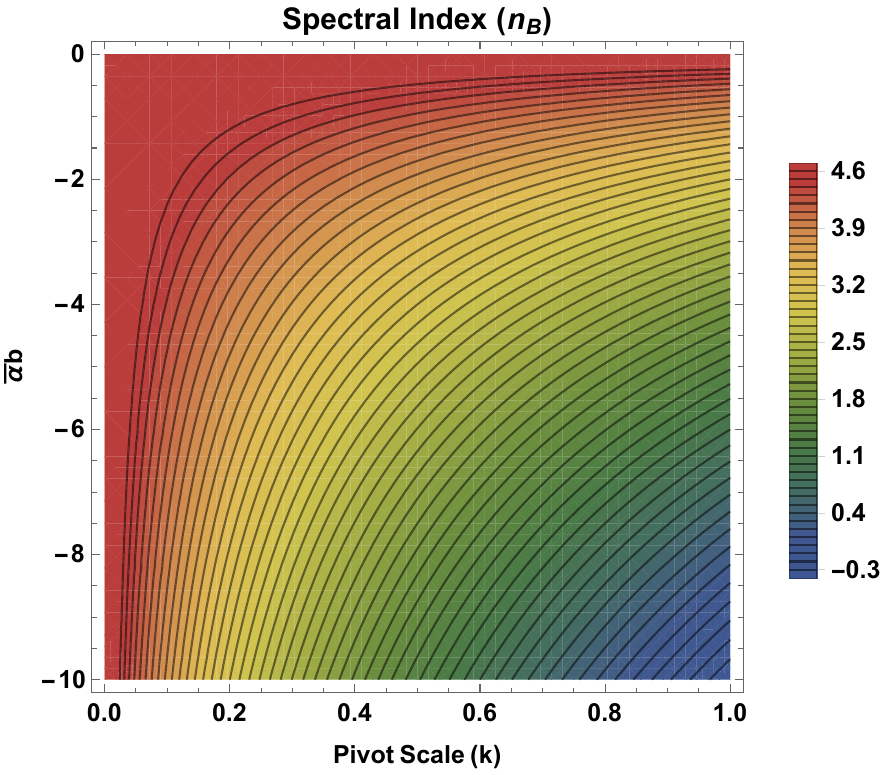}

\caption{The above figure demonstrates the magnetic spectral index $n_{B}$ with both the pivot scale $k$ (in Mpc$^{-1}$) and the parameter $\bar{\alpha}b$. Each contour depicts one constant value of the magnetic spectral index. As evident from the colour coding the spectral index varies from values greater than unity to values less than unity, while passing through one, which would depict scale invariance power spectrum.}\label{fig_04}
\end{center}
\end{figure*}

Keeping this in mind we have represented numerical estimations of the magnetic field strength for various choices of the parameter $\bar{\alpha}b$ for different pivot scales of measurement $k$ in \ref{Table_01}. As one can see from the table, given appropriate choices of the parameters, the magnetic field strength can range from $\sim 10^{-13}$ Gauss to $\sim 10^{-16}$ Gauss respectively. For larger values of $\bar{\alpha}b$, the strength of magnetic field also increases. However, for large values of $|\bar{\alpha}b|$ back-reaction of the electromagnetic field becomes important. On the other hand, in order to generate magnetic fields of appreciable strength today one may have to choose the combination $\bar{\alpha}b$ to be large enough. As evident from the previous discussion, this in turn will make the back-reaction effects important. Thus, as discussed in \cite{Durrer:2010mq}, in the case of a helical magnetic field there is indeed a certain tension between ignoring the back-reaction effect and generation of significant magnetic field. 

This can be appreciated further by exploring the situation associated with the electric field. From \ref{Mag_v2_35a} and \ref{Mag_v2_36a}, it immediately follows that the energy density of the electric field varies as $(-k\eta)^{-2}$ with respect to the magnetic field and hence can approach high values at the end of the inflation, when $k\eta \to 0$. Even though a strong electric field by itself is not problematic, since at the end stages of inflation it depletes quickly by the Schwinger pair-production mechanism, it is important to check whether the back-reaction of the electric field can still be neglected. The energy density of the electric field can be very easily determined by integrating $k^{-1}(d\rho_{\rm E}/d\ln k)$ over the scales of physical interest. If one assumes $\nu$ to be independent of the pivot scale $k$, the above integral can be explicitly computed and one will be able to infer that back reaction effects are negligible. However in the realistic scenario this is not the case and one must take into account the dependence of $\nu$ on the pivot scale k. This cannot be performed analytically due to the complicated dependence of $(d\rho _{\rm E}/d\ln k)$ on the pivot scale. Therefore we have computed it numerically and have presented the ratio $(\rho_{\rm E}/\rho_{\rm inf})$ for different choices of $|\bar{\alpha}b|$ in \ref{Table_02}, where $\rho_{\rm inf}$ corresponds to the energy scale of inflation. As evident from the computation, as $|\bar{\alpha}b|$ increases the energy density of the electric field also increases. If we want to keep the effect of back reaction negligible at the level of $10^{-10}$, one must have $|\bar{\alpha}b|<4$. It is clear from \ref{Table_01}, that the associated magnetic field strength in the present day universe can still become $\sim 10^{-13}~\textrm{Gauss}$, in accord with the observations.

\begin{table*}
\begin{center}
\caption{The numerical estimations of the ratio $\rho_{\rm E}/\rho_{\rm inf}$ is being presented for different choices of the parameter $\bar{\alpha}b$, where $\rho_{\rm E}$ being the energy density in the electric field and $\rho_{\rm inf}$ being the energy density of the inflationary field. As evident for smaller values of the parameter $|\bar{\alpha}b|$ the ratio is smaller and hence back reaction effects are small.}
\label{Table_02}       
%
%
\begin{tabular}{p{0.5cm}p{2.5cm}p{2.5cm}p{2.5cm}p{2.5cm}}
\hline\noalign{\smallskip}
\hline\noalign{\smallskip}
$|\bar{\alpha}b|$   &  ($\rho_{\rm E}/\rho_{\rm inf}$)  \\
\hline \noalign{\smallskip}
\hline \noalign{\smallskip}
0.1			    	& $4.621 \times 10^{-15}$\\
0.5				& $1.126 \times 10^{-13}$\\
1				& $4.477 \times 10^{-13}$\\
2				& $2.193 \times 10^{-12}$\\
\noalign{\smallskip}
\hline\noalign{\smallskip}
\hline \noalign{\smallskip}
\end{tabular}
\end{center}
\end{table*}

One can verify the above result from a different perspective as well. In the above analysis, we have discussed the total energy density in the electric field, by integrating over the pivot scale $k$. However one can also compare the maxima of the power spectrum in the electric field with the power spectrum of the inflaton field. In particular, the maxima in the power spectrum of the electric field appears at a certain value of the pivot scale $k_{\rm max}$, which satisfies the following algebraic equation, 
\begin{align}
k\ln k=\frac{\nu(k)\left\{3-2\nu(k)\right\}}{-\bar{\alpha}b}
\end{align}
As evident from the above expression, if $\nu$ was independent of the pivot scale, then the choice $\nu=(3/2)$ would make the power spectrum scale invariant. However, due to non-zero $b$ and $\bar{\alpha}$, it immediately follows that the theory has an inbuilt length scale $\bar{\alpha}b$ and hence this justifies the existence of a maximum, i.e., a special length scale in the problem. The energy density per unit pivot scale associated with the electric field at the maximum can thus be easily determined for a given $\bar{\alpha}b$. As mentioned earlier for larger values of $\bar{\alpha}b$, the electric field strength at the maximum will be comparable, or even larger than the energy density in the inflaton field. For example, with $|\bar{\alpha}b|\sim -5$, we have electric field strength at the maxima to be $\sim 10^{-3}$ times the energy density of the inflaton field. While for $|\bar{\alpha}b|\sim 0.5$, we have the electric field strength at the maximum to be $10^{-8}$ times the energy density of the inflaton field. Hence for smaller values of $|\bar{\alpha}b|$ the maximum of the energy density of the electric field is small compared to the energy density of the inflationary field at that scale. Hence the total energy density is also smaller, as depicted in \ref{Table_02}. 

Further as emphasized earlier, the magnetic field presented here shows scale dependance, as for the same choice of $\bar{\alpha}b$, for different pivot scales, the strength varies significantly. This feature is presented more robustly in \ref{fig_03}, where we have drawn various contours of constant magnetic field strength in the $(\bar{\alpha}b,k)$ plane. The colour coding demonstrates the strength of the magnetic field and shows that for smaller values of $\bar{\alpha}b$ and $k$ the strength is weak. Further, we have also investigated how the parameter $n_{B}$ ($n_B$ being the magnetic spectral index) varies with different pivot scales $k$ and the parameter $\bar{\alpha}b$ in \ref{fig_04}. For various choices of  $k$ and the parameter $\bar{\alpha}b$, the parameter $n_{B}$ goes from values larger than unity to values smaller than one. Reminiscent to scalar spectral index, any deviation of this parameter from unity essentially indicates scale-dependence of the power spectrum under consideration. So, this plot  gives rise to a (non-exhaustive) parameter space where one can have significant deviation from scale-invariant magnetic power spectrum. In particular, we can get both blue and red spectral tilt. As pointed out earlier, this essentially means that, along with magnetic power spectrum, one needs to estimate and constrain one more parameter, namely, magnetic spectral index, in order to get a hold on primordial magnetic field. Until now we have very feeble constraint on this, as analysed here using the lower bound of magnetic field strength as expected from CMB.  We do feel this feature is going to play a crucial role in constraining models for primordial magnetic field with upcoming CMB observations.  Thus we can get very interesting power spectrum and appreciable magnetic field strength along with appropriate magnetic spectral index for a wide parameter range in the model under consideration. 

Even though the back-reaction of the electromagnetic fields on the cosmological background can be neglected, one also needs to take into account the back-reaction of the scalar field. This needs to be handled carefully, since the rate of change of scalar field with conformal time goes as $a(\eta)$ and can become very large at the end of inflation. Thus one either needs some tailored initial condition or, one must take into account the back-reaction of the scalar field on the background geometry. However, the scale factor will definitely not remain exponential at the later part of the inflationary scenario and hence one needs to write down the evolution equations properly to address the issue of back-reaction of the scalar field on the background cosmology, this we leave for the future.
\section{Conclusions}

Generation of large scale magnetic field of appreciable strength in the present universe is a challenge. The most promising origin of such a magnetic field is primordial in nature. However, it turns out that due to conformal invariance, Maxwell's theory cannot account for the current strength of the large scale magnetic field. So far several models have been proposed which break the scale invariance or introduce additional helical modes of the electromagnetic theory. Unfortunately most of these models remained adhoc without much justification to the choice of various terms in the Lagrangian. In this work, we have proposed a new model which inhibits a term in addition to the Maxwell's action, which originates naturally from anomaly cancellation of the $U(1)$ gauge theory. This additional piece in the electromagnetic Lagrangian is absolutely essential to maintain gauge invariance of the electromagnetic Lagrangian at the quantum level. However, classically this term is sufficient to warrant that conformal invariance is kept intact while helical modes are generated in a hilltop inflationary scenario. This raises the hope of arriving at a model of primordial magnetogenesis, which has its origin from a fundamental picture. Starting from the associated action, we have derived the matter energy momentum tensor originating from the same and hence the modified Maxwell's equation. Due to explicit breaking of parity symmetry, the field equations satisfied by the vector potential in Fourier space inherits an additional contribution depending on the time derivative of a scalar field appearing naturally in the theory. For the time variation of the scalar field, fixed by the hilltop inflation model under consideration, we could arrive at the power spectrum of the electric and magnetic field, which is intrinsically scale dependent. It turns out that the scale dependance of the magnetic power spectrum is an important feature of this theory. Further, for reasonable values of  the parameter $\bar{\alpha}b$ (originating from the action) for different pivot scale of observation $k$, we could generate a magnetic field having appreciable strength. By choosing the parameters appropriately we have been able to generate a magnetic field strength with variations ranging from $10^{-13}$ Gauss to $10^{-16}$ Gauss. This makes the model robust enough by generating a magnetic field strength of desired amount along with concrete observational ramification in terms of scale dependent power spectrum with a fundamental origin.We leave the study of this model involving modified gravity theories for the future.
\section*{Acknowledgements}

Research of SC is funded by the INSPIRE faculty fellowship from the DST, Government of India (Reg. No. DST/INSPIRE/04/2018/000893)  and by the Start-Up Research Grant from SERB-DST, Government of India (Reg. No. SRG/2020/000409). Research of SSG is partially supported by the SERB-Extra Mural Research Grant (Reg. No. EMR/2017/001372), Government of India and SP thanks SERB-DST, Government of India for Research Grant No. NMICPS/006/MD/2020-21.
\bibliography{References}

\bibliographystyle{./utphys1}
\end{document}